\documentclass[11pt]{article}

\usepackage[a4paper,margin=1in]{geometry}
\usepackage[T1]{fontenc}
\usepackage[utf8]{inputenc}
\usepackage{lmodern}
\usepackage{microtype}
\usepackage{amsmath,amssymb,bm}
\usepackage{booktabs,array,longtable,tabularx}
\usepackage{enumitem}
\usepackage[nocompress]{cite}
\usepackage{hyperref}
\usepackage{xcolor}
\usepackage{setspace}
\usepackage{titlesec}
\usepackage{caption}
\usepackage{ragged2e}
\usepackage{graphicx}

\hypersetup{
  colorlinks=true,
  linkcolor=blue!50!black,
  citecolor=blue!50!black,
  urlcolor=blue!50!black,
  pdftitle={Experimental predictions of the E8 x omega E8 octonionic unification program},
  pdfauthor={Tejinder P. Singh}
}

\titleformat{\section}{\large\bfseries}{\thesection.}{0.6em}{}
\titleformat{\subsection}{\normalsize\bfseries}{\thesubsection.}{0.5em}{}
\setlist[itemize]{leftmargin=1.5em}
\setlist[enumerate]{leftmargin=1.6em}
\newcolumntype{Y}{>{\RaggedRight\arraybackslash}X}
\newcolumntype{L}[1]{>{\RaggedRight\arraybackslash}p{#1}}
\newcommand{\eights}{\texorpdfstring{$E_{8}\times \omega E_{8}$}{E8 x omega E8}}
\newcommand{\alphas}{\ensuremath{\alpha_s}}
\newcommand{\alphaem}{\ensuremath{\alpha_{\rm em}}}
\newcommand{\swt}{\ensuremath{\sin^2\theta_W}}
\newcommand{\MZ}{\ensuremath{M_Z}}
\newcommand{\SM}{Standard Model}
\newcommand{\ew}{electroweak}
\newcommand{\MSbar}{\overline{\rm MS}}

\begin{document}

\begin{center}
{\LARGE \textbf{Experimental predictions of the \eights\ octonionic unification program}}\\[0.35em]
{\large A falsification-oriented catalogue for quantum foundations, particle physics, gravitation, and cosmology}\\[0.8em]
{\large Tejinder P. Singh}\\[0.4em]
\textit{Tata Institute of Fundamental Research, Mumbai 400005, India}\\
July 4, 2026 (v2; v1: April 7, 2026)
\end{center}

\begin{abstract}
\noindent The \eights\ octonionic program aims at a deliberately ambitious synthesis: quantum theory without external classical time, objective collapse, emergent classical space-time, exceptional Jordan-algebra flavor structure, and an exceptional-group unification of visible matter with a right-handed pre-gravitational sector. The purpose of the present paper is not to review the whole formalism, but to assemble in one place the claims that are experimentally vulnerable and to classify them by logical strength. We begin with a cold-start pedagogical map from the core ingredients of the program to the observables it claims to generate. This map matters because the program's breadth is both its attraction and its principal vulnerability: if the particle-physics, gravitation, and quantum-foundational claims are not visibly derived from the same structure, the framework reduces to a collection of disconnected conjectures.

On the quantum-foundational side the program predicts objective spontaneous collapse, operator time, spontaneous collapse in time, loss of temporal interference above an attosecond-scale separation, a six-dimensional explanation of apparent nonlocality, possible Bell correlations beyond the Tsirelson bound, a fermion-only collapse sector, and holographic or Karolyhazy-type space-time uncertainty. In particle physics it predicts a right-handed pre-gravitational gauge sector, an extended Higgs sector, dark electromagnetism and its dark photon, three right-handed Majorana neutrinos --- now assigned a cosmological role as diluted $\simeq\!40$\,eV relics that close the matter budget --- Majorana light neutrinos, inverted neutrino mass ordering with $m_{\beta\beta}\simeq18$\,meV, leptonic Dirac CP conservation ($\delta_{CP}^\ell\in\{0,\pi\}$, conditional on a computable Higgs-bridge term; this corrects the maximal-phase entry of v1), CKM root-sum rules, charged-fermion mass relations including the first-generation $1{:}4{:}9$ pattern and the relation $m_\tau/m_\mu = m_s/m_d$, a low-energy fine-structure constant, a weak-mixing-angle derivation, and the mixed-regime relation $\alphas(\MZ)/\alphaem(0)=16$. In gravitation and cosmology it predicts parity-violating $SU(2)_R$ pre-gravitation, a gravi-weak relation in split-signature six dimensions, emergent classical gravity rather than a fundamental graviton, a relativistic MOND-like infrared regime with $a_0$ anchored to $\Lambda$, a diluted dark thermal sector with a two-epoch dark-radiation fingerprint ($\Delta N_{\rm eff}\simeq0.19$ at nucleosynthesis falling to $\simeq0.05$ at recombination), and permanent null results in all particle-dark-matter searches.

The paper is written as a catalogue of possible failure modes rather than a success list. We therefore distinguish distinctive predictions from generic beyond-the-\SM\ features, include a quantitative status table for the sharpest claims, state openly where present data already create tension, and identify the three most plausible routes to a truly discriminating test: a concrete Bell protocol yielding $\mathrm{CHSH}>2\sqrt{2}$, a quantitative fermion-versus-boson collapse test, and a fully scheme-consistent global \ew-scale analysis of the program's coupled flavor and gauge relations. Version 2 corrects the v1 leptonic-phase entry, updates the sterile-neutrino sector to its cosmological role, and adds a consolidated predictions ledger (Table~\ref{tab:ledger}): deciding experiment, timeline, kill condition, and commitment date for each sharp claim.
\end{abstract}

\noindent\textbf{Keywords:} octonions; exceptional Jordan algebra; \eights; spontaneous collapse; Majorana neutrinos; dark electromagnetism; relativistic MOND; Bell nonlocality; gauge couplings; fermion mass ratios

\section{Introduction and scope}

The \eights\ program did not begin as a conventional beyond-the-\SM\ proposal. Its starting point was the claim that ordinary quantum theory presupposes an external classical time, even though a fully quantum  universe should not. That foundational concern was coupled to a second one: the measurement problem suggests that strictly unitary evolution cannot be the whole microscopic story \cite{Singh2018QTSTS,Singh2018GRWST,Singh2019STCollapse}. In the post-2018 literature, these themes are developed through generalized trace dynamics in the sense of Adler \cite{Adler2004}, operator-valued space-time, spontaneous localization, division algebras, the exceptional Jordan algebra, and finally an explicit exceptional-group unification picture \cite{Singh2021Alpha, Singh:2022lpw,
Kaushik2022E8,Singh2025SplitBioctonions,Singh2026Emergent}. The result is a framework that makes, or at least aspires to make, testable claims across quantum foundations, flavor physics, gauge structure, gravitation, and cosmology.

This paper has a deliberately narrow purpose. It is not a full review of the formalism, and it is not an advocacy piece. Its task is to identify what the program currently claims to predict, how those predictions are supposed to arise, which of them are distinctive to this framework, which of them are generic beyond-the-\SM\ features, and where present data already press back. A program this broad either wins credibility by exposing itself to cross-sector falsification, or loses credibility by sounding like a branded list of unrelated possibilities.

A central difficulty is that the program still does not possess a universally recognized flagship test analogous to the 1919 light-bending check of general relativity. Yet the present literature has already come close to three candidate discriminants. The first is a genuine violation of the Tsirelson bound in Bell experiments \cite{Ahmed2022Tsirelson,Furquan2025Timelike,Cirelson1980,PopescuRohrlich1994}. The second is the claim that intrinsic objective collapse acts on fermions but not on bosons \cite{Singh2026FermionCollapse}. The third is a dense overconstraint of flavor and gauge relations that the \SM\ does not predict at all \cite{Singh2025MassRatios,Singh2026Gauge}. This paper asks, as soberly as possible, how close any of these routes is to becoming truly decisive.

The broader mathematical context should also be stated openly. Octonions, Jordan algebras, and exceptional groups have long appeared in attempts to understand quark and lepton structure \cite{Jordan1934,GunaydinGursey1973,GunaydinGursey1974,Baez2002}. A modern literature connects complex octonions, Clifford algebras, exceptional Jordan geometry, and related algebraic structures to generation structure and \SM\ representations \cite{Furey2014,Furey2018,GillardGresnigt2019,Todorov2023}. What is distinctive here is not the bare use of octonions, but the attempt to weld them to trace dynamics, objective collapse, emergent space-time, and an explicit visible/pre-gravitational exceptional branching.

\section{From formal ingredients to observables: a cold-start map}

A persistent criticism of ambitious unification programs is that a new reader cannot tell how the predictions are derived. That criticism is fair here unless the logical dependencies are made explicit. The basic claim of the \eights\ program is not that one starts from a single equation and reads off all observed physics. The claim is instead that several tightly linked structural ingredients generate several families of observables. Table~\ref{tab:flow} is a compact map.

A useful way to state the deeper symmetry principle is that the pre-geometric laws are covariant under automorphisms of the octonionic/Jordan algebraic structure from which the \eights\ framework is built.  Since $\mathrm{Aut}(\mathbb{O})\cong G_2$, octonion automorphisms should be viewed as the seed algebraic symmetry, not as the whole low-energy symmetry group; in the exceptional Jordan and \eights\ setting this seed is enlarged by triality, Jordan-algebraic structure, and exceptional branching into the broader framework from which the phenomenology is extracted \cite{Baez2002,Kaushik2022E8,Bhatt2022Majorana,Patel2023CKM,Singh2025SplitBioctonions,Singh2025MassRatios}.

What symmetry breaking does in this program is therefore more subtle than turning octonion automorphisms directly into space-time diffeomorphisms. Rather, before classicalization there is a single pre-spatiotemporal algebraic arena in which the distinction between ``internal'' and ``external'' symmetry has not yet emerged. Spontaneous localization yields an emergent classical space-time, while concurrent triality and electroweak breaking select a definite vacuum and separate the residual internal gauge symmetries from the diffeomorphism invariance of the emergent manifold. In that precise sense, the program aims to unify gauge symmetry and space-time symmetry: not by asserting a literal group-theoretic identity $\mathrm{Aut}(\mathbb{O}) \simeq \mathrm{Diff}(M)$, which would be false, but by treating both as low-energy descendants of one deeper exceptional-octonionic algebraic structure \cite{Singh2018QTSTS,Singh2018GRWST,Singh2019STCollapse,Kaushik2022E8,Singh2023GraviWeak,Singh2025SplitBioctonions,Wesley2026SO33,Singh2025MassRatios}.

\begin{table}[ht]
\centering
\caption{Cold-start derivation map: how the main ingredients of the program are claimed to feed into observables.}
\label{tab:flow}
\begin{tabularx}{\textwidth}{L{0.27\textwidth} L{0.33\textwidth} Y}
\toprule
Formal ingredient & Derived structural claim & Claimed observables \\
\midrule
Generalized trace dynamics with operator-valued space-time \cite{Adler2004,Singh2018QTSTS,Singh2018GRWST,Singh2019STCollapse} & Anti-self-adjoint fluctuations drive objective collapse; classical space-time emerges only after localization & Mesoscopic collapse phenomenology; anomalous heating; temporal interference and collapse in time; emergent classical metric sector \\
Exceptional Jordan algebra $J_3(\mathbb{O}_\mathbb{C})$, triality, and flavor ladders \cite{Bhatt2022Majorana,Patel2023CKM,Singh2025MassRatios,TeliSinghCP2026} & Three generations from Jordan eigenvalues; fixed ladder geometry and Dynkin swap; concurrent \ew/triality breaking & Charged-fermion mass relations, Koide shift, CKM root-sum rules, Majorana neutrinos, leptonic CP structure localized by transport theorems \\
Exceptional branching and split bioctonionic geometry \cite{Kaushik2022E8,Singh2025SplitBioctonions} & One exceptional factor feeds the visible sector; the second feeds a right-handed pre-gravitational sector; two extra $SU(3)$ factors support a $(3,3)$ backbone & Additional Higgs structure, right-handed gauge sector, sterile or inert right-handed states, dark photon, six-dimensional causal reinterpretation of nonlocality \\
Gravi-weak construction and right-handed pre-gravitation \cite{Singh2023GraviWeak,Wesley2026SO33} & Weak and gravitational sectors share a broken higher-dimensional origin; parity violation is a relic of a deeper right-handed asymmetry & Parity-violating $SU(2)_R$ pre-gravitation, weak-scale estimates, split-signature gravi-weak phenomenology \\
Dark electromagnetism sourced by square-root mass \cite{Finster2024DarkEM,Singh2026RelMOND,SinghCosmo2026} & Infrared long-range force tied to the same right-handed sector that appears in particle physics & Relativistic MOND-like regime with $a_0$ anchored to $\Lambda$, galaxy-dynamics modifications, lensing, and a diluted dark thermal sector with a two-epoch $\Delta N_{\rm eff}$ fingerprint \\
Electroweak/triality concurrency \cite{Singh2023GraviWeak,Singh2025MassRatios,Singh2026Gauge} & The practically relevant matching scale for flavor, pre-gravitation, and visible couplings is shifted from the Planck scale to the \ew\ scale & Common-scale flavor tests near \MZ; weak-scale rather than Planck-scale quantum-gravity imprint in the program's phenomenology \\
\bottomrule
\end{tabularx}
\end{table}

Two comments are essential. First, the program is not claiming that all quantum-gravity effects become order-one at $100\,\mathrm{GeV}$. Its claim is subtler: the operative matching point at which flavor structure, pre-gravitation, and visible-sector physics become tied to one another is reset from the Planck scale to the \ew\ scale. In that sense the framework seeks to unify the \SM\ with gravity at the \ew\ scale, and its sharpest flavor and coupling tests are therefore supposed to be evaluated near \MZ\ rather than only in a far-ultraviolet extrapolation \cite{Singh2023GraviWeak,Singh2025MassRatios,Singh2026Gauge}.

Second, this ``reset'' of the practically relevant quantum-gravity scale is a claim of the framework, not an established empirical fact. It is useful only if it sharpens comparisons rather than blurring them. One immediate consequence is methodological: if the theory says that triality breaking and \ew\ breaking are concurrent, then a fair test of the mass and coupling relations must be performed at a common renormalization scale near the \ew\ threshold \cite{HuangZhou2021,Singh2025MassRatios}. That raises the bar for both proponents and critics.

\section{What counts as a prediction?}

A framework spanning quantum foundations, particle phenomenology, gravitation, and cosmology will inevitably make claims of unequal logical strength. To avoid overstatement, it is useful to sort them before discussing them in detail.

\begin{table}[ht]
\centering
\caption{Predictive classes used in this review.}
\label{tab:classes}
\begin{tabularx}{\textwidth}{L{0.10\textwidth} L{0.30\textwidth} L{0.28\textwidth} Y}
\toprule
Class & Meaning & Representative examples & Typical failure mode \\
\midrule
A & Parameter-free quantitative relation & $\sqrt{m_e}:\sqrt{m_u}:\sqrt{m_d}=1:2:3$; $m_\tau/m_\mu = m_s/m_d$; low-energy $\alpha$; $\alphas(\MZ)/\alphaem(0)=16$ & Same-scheme, same-scale comparison remains inconsistent after stated matching procedure \\
B & Semi-quantitative scale or regime estimate & Temporal-interference cutoff near $10^2$ as; weak-scale estimate; MOND-like regime below an acceleration threshold & Observed scale or regime lies outside the claimed window \\
C & Structural beyond-the-\SM\ prediction & Right-handed gauge sector; extra Higgs; dark photon; inert right-handed Majorana states; split-signature gravi-weak backbone & Dedicated searches or consistency checks eliminate the required sector \\
D & Qualitative beyond-standard signature & CHSH above $2\sqrt{2}$; fermion-only collapse; apparent nonlocality explained via extra timelike dimensions & Experiments continue to agree with standard quantum theory in the only regime where the theory says they should fail \\
\bottomrule
\end{tabularx}
\end{table}

This classification should be supplemented by a second distinction: some predictions are distinctive to the \eights\ program, whereas others are generic beyond-the-\SM\ features that also occur in many unrelated models. A second Higgs boson, sterile neutrinos, and a dark photon are important structural claims, but none is unique. By contrast, spontaneous collapse in time, a six-dimensional causal reinterpretation of Bell nonlocality, a possible Bell-window violation above Tsirelson, the relation $m_\tau/m_\mu = m_s/m_d$ arising from a Dynkin swap, and the mixed-regime coupling relation $\alphas(\MZ)/\alphaem(0)=16$ are much more distinctive. This distinction matters because it determines which data could discriminate the program from generic model-building rather than merely support it.

\section{Consolidated catalogue of present empirical claims}

Table~\ref{tab:catalogue} assembles the principal claims presently extractable from the post-2018 literature. It is intentionally more discriminating than a simple list: the table flags whether a claim is distinctive to this framework or generic across broader beyond-the-\SM\ model space.

{\small
\begin{longtable}{@{}L{0.14\textwidth} L{0.27\textwidth} L{0.12\textwidth} L{0.27\textwidth} L{0.12\textwidth}@{}}
\caption{Catalogue of the main empirical claims presently associated with the \eights\ program.}\label{tab:catalogue}\\
\toprule
Sector & Claim & Specificity & Best current handle & Representative sources \\
\midrule
\endfirsthead
\toprule
Sector & Claim & Specificity & Best current handle & Representative sources \\
\midrule
\endhead
Quantum foundations & Objective spontaneous collapse and emergent classical space-time & Moderate & Mesoscopic interferometry; force-noise, spontaneous-radiation, and heating constraints & \cite{Singh2018GRWST,Singh2019STCollapse,Mishra2018BulkHeat,GRW1986,Pearle1989,Bassi2013,Bassi2023,Aprile2026XENON} \\
Quantum foundations & Operator time, spontaneous collapse in time, and loss of temporal interference for separations above $\mathcal{O}(10^2\,\mathrm{as})$ & High & Time-domain double-slit experiments with controlled attosecond separation & \cite{Singh2019STCollapse,Lindner2005,Kaneyasu2023} \\
Quantum foundations & Extra timelike dimensions explain apparent nonlocality & High & Bell tests with explicit timing geometry and causal reconstruction in six dimensions & \cite{Ahmed2022Tsirelson,Furquan2025Timelike,JavedWilsonEwing2026} \\
Quantum foundations & Possible CHSH violation above the Tsirelson bound: mechanism and sign (positive excess) supplied, magnitude \emph{not yet derived} --- the dedicated 2026 assessment declares the prediction not yet falsifiable until it is; two motivated regimes identified (short elapsed proper time between pair creation and basis choice; centre-of-mass energies above the electroweak scale) & Very high if a magnitude is derived; currently a mechanism with a search frontier & Most precise CHSH: $S=2.82759\pm0.00051$, $1.6\sigma$ below $2\sqrt2$, headroom $\lesssim\text{few}\times10^{-4}$; proposed: tomography-anchored photonic null test at $10^{-5}$, time-resolved $S(\tau)$ to sub-picosecond, mass-and-complexity ladder, $t\bar t$/$\tau^+\tau^-$/hyperon spin correlations & \cite{SinghTsirelson2026,Ahmed2022Tsirelson,Furquan2025Timelike,Cirelson1980,PopescuRohrlich1994,Hensen2015,Giustina2015,Shalm2015} \\
Quantum foundations & Fermion-only collapse, with no intrinsic bosonic collapse & Very high & Comparative mesoscopic superposition or heating experiments isolating fermion-rich and boson-dominated sectors & \cite{Singh2026FermionCollapse} \\
Quantum foundations & Holographic or Karolyhazy uncertainty and minimum length & Moderate & Interferometric quantum-geometry-noise searches and precision metrology & \cite{Singh2021Holography,Karolyhazy1966,NgVanDam1994,Chou2017,Richardson2021} \\
Particle physics & Extended Higgs sector: a complete second electroweak doublet --- physical quartet $(H, A, H^\pm)$ with fixed charges, CP-conserving 2HDM near alignment, Type-I-like Yukawas; mass not determined, with the electroweak-matching claim motivating (not forcing) $10^2$--$10^3$\,GeV & Structural (Class C), now phenomenologically profiled & Alignment-robust electroweak pair production ($H^+H^-$, $H^\pm A$, $H^\pm H$, $HA$); Higgs-coupling precision at HL-LHC and FCC-ee; LEP charged-Higgs, EWPT and flavor constraints & \cite{Kaushik2022E8,SinghHiggs2026,PDG2024} \\
Particle physics & Right-handed pre-gravitational gauge sector $SU(3)_{\mathrm{grav}}\times SU(2)_R\times U(1)_g$ & Moderately distinctive & Indirect low-energy signatures; consistency with \ew\ and astrophysical constraints & \cite{Kaushik2022E8,Singh2023GraviWeak,Wesley2026SO33} \\
Particle physics & Dark photon associated with dark electromagnetism & Generic BSM in existence, distinctive in origin & Dark-sector searches and astrophysical limits & \cite{Kaushik2022E8,Finster2024DarkEM,PDG2024} \\
Particle physics & Three right-handed Majorana neutrinos, realised as diluted thermal relics of common mass $\simeq\!40$\,eV that close the matter budget ($\mathrm{tr}\,M_R\simeq113$\,eV fixed by $\Omega_m$; each mass free within $\sim\![17,100]$\,eV) & Highly distinctive in role & Cosmology: two-epoch $\Delta N_{\rm eff}$ pattern and the free-streaming split (clusters clustered, galaxies bare); KATRIN-type kink searches predicted null, $|U_{e4}|^2\lesssim10^{-6}$ & \cite{Vaibhav2023Sterile,Singh2025MassRatios,SinghCosmo2026,SinghNu2026,PDG2024} \\
Particle physics & Majorana nature of light neutrinos & Generic in BSM, stronger here as an internal no-go for Dirac neutrinos & Neutrinoless double-beta decay: $m_{\beta\beta}=18.2\pm1.2$\,meV in the inverted-ordering package (opposite pair parities, lower edge of the IO band) & \cite{Bhatt2022Majorana,Singh2025MassRatios,TeliSinghCP2026,SinghNu2026,KamLANDZen2023,LEGEND2026} \\
Particle physics & Leptonic Dirac CP conservation, $\delta_{CP}^\ell\in\{0,\pi\}$, unless a specific identity--flavor Higgs-bridge component is nonzero (corrects the v1 maximal-phase entry) & Distinctive and sharply localized & DUNE and Hyper-K; a measured phase would localize the bridge term rather than falsify & \cite{TeliSinghCP2026,SinghNu2026,NuFIT2024,DUNE2022,HyperK2026} \\
Particle physics & Inverted neutrino mass ordering, with $(m_1,m_2,m_3)=(49.07,49.84,0.762)$\,meV and $\Sigma m_\nu\simeq0.10$\,eV (minimal Majorana branch; the only branch of the combined construction with the cosmological sterile sector) & Highly distinctive as exposure: no normal-ordering fallback remains & JUNO ordering determination; cosmological $\Sigma m_\nu$ within the program's own background & \cite{TeliSinghCP2026,SinghNu2026,SinghCosmo2026,JUNO2024Order} \\
Particle physics & First-generation relation $\sqrt{m_e}:\sqrt{m_u}:\sqrt{m_d}=1:2:3$ & Highly distinctive & Common-scale running-mass analysis & \cite{Bhatt2022Majorana,Singh2025MassRatios,HuangZhou2021} \\
Particle physics & Cross-sector relation $m_\tau/m_\mu = m_s/m_d$ (equivalently $\sqrt{m_\tau/m_\mu}=\sqrt{m_s/m_d}$) & Highly distinctive & Common-scale running-mass analysis & \cite{Singh2025MassRatios,HuangZhou2021} \\
Particle physics & CKM root-sum rules and leading small-entry predictions & Highly distinctive & Global CKM fits with common-scale inputs & \cite{Patel2023CKM,Singh2025MassRatios,PDG2024} \\
Particle physics & Exact Koide in the symmetric phase and a definite shifted Koide relation after breaking & Highly distinctive & Common-scale flavor analysis & \cite{Singh2025MassRatios,Koide1983} \\
Particle physics & Low-energy fine-structure constant from octonionic or Jordan structure & Distinctive & Comparison with CODATA low-energy $\alpha$ & \cite{Singh2021Alpha,Mohr2025} \\
Particle physics & Weak-mixing-angle derivation with $\swt\simeq 0.25$ & Distinctive but presently strained & Precision \ew\ comparison with explicit convention matching & \cite{Raj2022Bosonic,LHCb2024WeakAngle} \\
Particle physics & Mixed-regime coupling relation $\alphas(\MZ)/\alphaem(0)=16$ & Highly distinctive & Comparison to measured $\alphas$ and $\alpha$ with explicit statement of mixed-scale convention & \cite{Singh2026Gauge,PDG2024,Mohr2025} \\
Particle physics / gravity & Universally interacting spin-1 Lorentz bosons rather than a fundamental graviton in the pre-quantum description & Highly distinctive & Structural consequence; indirect only at present & \cite{Singh2021Alpha,Singh2026Emergent} \\
Gravity/cosmo & Dark electromagnetism as a new long-range force sourced by square-root mass & Highly distinctive & Galaxy dynamics, lensing, and cosmological phenomenology & \cite{Finster2024DarkEM,Singh2026RelMOND} \\
Gravity/cosmo & Relativistic MOND-like infrared regime with GR recovered at high acceleration & Moderately distinctive in mechanism, not in phenomenology & Rotation curves, wide binaries, clusters, lensing, CMB and large-scale structure & \cite{Finster2024DarkEM,Singh2026RelMOND,SinghCosmo2026,Milgrom1983,Bekenstein2004,SkordisZlosnik2021} \\
Gravity/cosmo & Two-epoch dark radiation: $\Delta N_{\rm eff}^{\rm BBN}\simeq0.19$ falling to $\Delta N_{\rm eff}^{\rm CMB}\simeq0.054$ & Highly distinctive pattern: a massless dark fluid gives equal values at both epochs, no dark radiation gives zero at both & BBN abundances (D/H, $Y_p$) versus CMB $N_{\rm eff}$ (Simons Observatory, CMB-S4) & \cite{SinghCosmo2026} \\
Gravity/cosmo & Parity-blindness of everyday gravity \emph{derived} from $SU(2)_R$ pre-gravitation: three screening theorems (soft-graviton consistency; no chirality for massive matter, $\gtrsim98\%$ of baryonic mass chirality-blind; single conformal structure from gluing); the coupling-level reading ``only right-handed matter gravitates'' independently excluded by SN1987A ($2\times10^{-3}$) and cosmology ($N_{\rm eff}$, CMB phase shift) & Derived consistency, plus a data-level exclusion the programme passes & Universality/free-fall tests; supernova neutrino--GW timing & \cite{SinghParity2026,Singh2023GraviWeak,Wesley2026SO33} \\
Gravity/cosmo & Residual gravitational parity violation confined to a named operator basket: Pontryagin and Nieh--Yan terms (GW catalogues bound $M_{\rm PV}<6.5\times10^{-42}$\,GeV $\simeq4.5\,H_0$ --- inside the Hubble-scale window the programme's Connes-time structure independently selects), a maximally chiral graviton vacuum at the self-dual point $\gamma=\pm i$ (CMB TB/EB correlations observable if $r\gtrsim\text{few}\times10^{-3}$), spin-gravity effects bounded at $10^{-21}$, ultra-soft $SU(2)_R\times U(1)_{\rm dem}$ vectors below $10^{-14}$ of gravitational strength & Highly distinctive; now quantitative & GW polarization catalogues; LiteBIRD/CMB-S4 TB/EB; laboratory spin-gravity; proposed Galactic-supernova triple-timing test ($\nu_e$, $\bar\nu_e$, GW) at the $3\times10^{-9}$ level & \cite{SinghParity2026} \\
Gravity/cosmo & Split-signature six-dimensional backbone with two overlapping Lorentzian sectors & Highly distinctive & Indirectly through Bell, parity, gravi-weak, and dark-sector signatures & \cite{Furquan2025Timelike,Kaushik2022E8,Singh2025SplitBioctonions,Wesley2026SO33} \\
Gravity/cosmo & Semi-quantitative estimate of the weak scale or Fermi constant from geometry & Distinctive but immature & Internal consistency and phenomenological sharpening still needed & \cite{Singh2023GraviWeak} \\
Gravity/cosmo & Emergent classical gravity rather than a fundamental quantum graviton & Highly distinctive as ontology, indirect as phenomenology & Structural consistency with the rest of the program & \cite{Singh2021Alpha,Singh2026Emergent} \\
\bottomrule
\end{longtable}}

\section{Quantum-foundational predictions}

\subsection{Objective collapse and operator space-time}

The foundational core of the program is the claim that ordinary unitary quantum theory is emergent rather than exact. Classical space and time are not primary; they arise only after spontaneous localization events in a deeper matrix-valued dynamics \cite{Singh2018QTSTS,Singh2018GRWST,Singh2019STCollapse}. Relative to standard GRW/CSL-type phenomenology \cite{GRW1986,Pearle1989,Bassi2013,Bassi2023}, the distinctive move is to make collapse responsible not only for the quantum-to-classical transition but also for the emergence of classical space-time itself.

That is already an empirical commitment. Any sufficiently successful large-mass interference experiment, or any stronger null bound on collapse-induced heating, force noise, or spontaneous radiation, constrains the program just as it constrains other collapse models. The \eights\ program is therefore falsifiable at the level of foundational phenomenology even before one invokes its exceptional-algebraic unification structure.

\subsection{Interference in time and spontaneous collapse in time}

One of the most attractive predictions in the literature is the temporal analogue of the spatial double slit. If time is an operator prior to classicalization, then one should expect temporal interference. But the program does not predict unlimited temporal coherence. It predicts spontaneous collapse in time as well as in space, and this is precisely why temporal interference should disappear once the time-slit separation becomes too large \cite{Singh2019STCollapse}. The estimate advocated in the operator-time paper is that interference should not survive for separations significantly larger than about $10^2$ attoseconds.

This is a real experimental target because attosecond time-domain double-slit physics already exists. Lindner \emph{et al.}\ demonstrated attosecond double-slit interference in photoionization, and Kaneyasu \emph{et al.}\ later realized time-domain double-slit control with synchrotron radiation \cite{Lindner2005,Kaneyasu2023}. These experiments were not designed to test operator time, but they show that the program's relevant scale is accessible. What is still missing is an experimentally usable visibility law: the literature contains the threshold estimate, but not yet a detailed, apparatus-level prediction for how the fringe contrast should fall with time-slit separation.

\subsection{Non-interferometric collapse signals}

The same collapse sector also predicts non-interferometric signatures, most notably anomalous bulk heating. Mishra, Vinante, and Singh estimated that underground low-background calorimetry could probe collapse-heating rates around $\lambda\sim 10^{-16}\,\mathrm{s}^{-1}$ \cite{Mishra2018BulkHeat}. This route remains valuable because it scales differently from matter-wave interferometry and is often experimentally cleaner \cite{Bassi2013,Bassi2023,Carlesso:2022pqr
}.

The experimental landscape has recently tightened. XENONnT has reported new world-leading bounds from a search for spontaneous X-ray emission, improving earlier Markovian CSL limits by about two orders of magnitude and arguing that the original white-noise CSL benchmark region is now experimentally excluded \cite{Aprile2026XENON}. That result does not directly falsify Singh's operator-space-time program, because the underlying collapse dynamics are not identical \cite{Kakade2023Spont} and coloured noise is the favoured theoretical possibility. But it changes the practical context: any new collapse proposal now enters an environment in which non-interferometric bounds are no longer merely aspirational.

\subsection{Extra timelike dimensions, apparent nonlocality, and the Tsirelson bound}

The boldest foundational claim is that our observed world is effectively embedded in six dimensions with two additional timelike directions, so that correlations appearing acausal in four dimensions can be causal in the deeper geometry \cite{Ahmed2022Tsirelson,Furquan2025Timelike}. This is already a substantial claim. But the program goes further: it suggests that Bell correlations may exceed the ordinary quantum Tsirelson limit $2\sqrt{2}$ in the relevant pre-quantum regime \cite{Ahmed2022Tsirelson,Furquan2025Timelike}.

This possibility is conceptually dramatic because it would falsify ordinary quantum mechanics itself, not merely local hidden-variable models. Bell's theorem and the CHSH inequality exclude local realism, while Tsirelson's theorem limits quantum violations to $2\sqrt{2}$ \cite{Bell1964,CHSH1969,Cirelson1980}. Popescu and Rohrlich emphasized that stronger-than-quantum yet no-signaling correlations are logically possible \cite{PopescuRohrlich1994}. The loophole-free Bell experiments of 2015 confirmed quantum nonlocality while remaining fully consistent with the Tsirelson limit \cite{Hensen2015,Giustina2015,Shalm2015}.

The right way to state the program's claim is therefore cautious, and the dedicated 2026 assessment \cite{SinghTsirelson2026} now does this with deliberate severity: the programme supplies a mechanism and a sign (a positive excess above $2\sqrt2$ in the non-equilibrated pre-quantum regime and in the six-dimensional bulk), but no derived magnitude, and the prediction is declared \emph{not yet falsifiable} until a magnitude is derived --- identified there as the single most valuable open theoretical task of this sector. The same paper confronts the record (the most precise CHSH measurement, $S=2.82759\pm0.00051$, sits $1.6\sigma$ below $2\sqrt2$, leaving room for a constant positive excess of at most a few $\times10^{-4}$; loophole-free tests across photons, atoms, NV centres and superconducting circuits show no anomaly), tabulates the benchmark suppression scalings against achievable precision, identifies the two regimes the programme itself motivates (short elapsed proper time; centre-of-mass energies above the electroweak scale --- the electroweak-threshold conjecture), and proposes five experiments, from a $10^{-5}$ tomography-anchored photonic null test and sub-picosecond time-resolved $S(\tau)$ to collider spin correlations. Until the magnitude is derived, this remains a \emph{candidate} flagship test with a well-posed search frontier, not a finished one.

There is also a broader context worth mentioning. Javed and Wilson-Ewing have recently shown that hydrogen spectroscopy can constrain a distinct geometric idea, namely wormhole-mediated entanglement in the ER=EPR spirit \cite{JavedWilsonEwing2026}. That mechanism is not the same as Singh's extra-time proposal, but the comparison is useful: both attempt to convert a geometrical account of entanglement into ordinary laboratory observables. The lesson is methodological. If a deep geometric account of nonlocality is right, it should eventually show up not only in conceptual reinterpretation but also in an experimentally calculable protocol.

\subsection{Fermion-only collapse}

A new 2026 claim is that the anti-self-adjoint term responsible for intrinsic collapse is present only in the fermionic sector and absent in the pure bosonic sector \cite{Singh2026FermionCollapse}. If correct, this would be one of the most distinctive predictions anywhere in the program. Standard quantum theory predicts no objective collapse for either bosons or fermions, while generic CSL models do not usually impose such a basic boson--fermion asymmetry.

At the same time, the present status should not be overstated. The current statement is qualitatively sharp but not yet a complete phenomenological prediction. Real mesoscopic systems are mixed and environmentally open; most interesting bodies contain both bosonic and fermionic degrees of freedom. What is needed next is a rate formula for realistic settings and an experimentally meaningful definition of what counts as a sufficiently bosonic or fermionic test system.

\subsection{Holographic uncertainty and minimum length}

A further foundational prediction is a Karolyhazy- or holography-type uncertainty relation,
\begin{equation}
(\delta \ell)^3 \sim L_P^2\,\ell,
\end{equation}
interpreted as a manifestation of quantum space-time \cite{Singh2021Holography,Karolyhazy1966,NgVanDam1994}. The point is not merely conceptual. Such a relation motivates interferometric and clock-based probes of space-time noise.

Here caution is essential. The Fermilab Holometer has already placed strong limits on several correlated quantum-geometry-noise models \cite{Chou2017,Richardson2021}. These exclusions do not automatically falsify the particular Karolyhazy relation invoked in Singh's work, because the measured observables are model dependent. The correct statement is therefore not that the terrain is untouched, but that a direct mapping from the specific uncertainty relation used in the \eights\ program to contemporary interferometric observables remains to be carried out.

\section{Particle-physics predictions}

\subsection{What is generic and what is distinctive?}

Before discussing details, it is worth separating the program's particle-physics claims into two bins. Generic beyond-the-\SM\ features include an extended Higgs sector, sterile or heavy neutral leptons, a dark photon, and a right-handed gauge sector. These are all interesting, but none uniquely identifies the framework. Distinctive claims are different: the flavor relations derived from the Jordan ladder, the Dynkin-swap relation $m_\tau/m_\mu = m_s/m_d$, the concurrency of triality breaking with \ew\ breaking, the mixed-regime coupling relation $\alphas(\MZ)/\alphaem(0)=16$, and the claim that the operative visible-sector/pre-gravitational matching scale is the \ew\ scale rather than the Planck scale.

That second bin is where the framework either earns or loses its phenomenological identity. If only the generic features survive, the program may remain a suggestive algebraic language but will not become a uniquely testable theory.

\subsection{Visible sector, right-handed sector, and the electroweak matching scale}

The 2022 unification paper proposes that one exceptional factor branches to the visible \SM\ sector while the second branches to a right-handed pre-gravitational sector $SU(3)_{\mathrm{grav}}\times SU(2)_R\times U(1)_g$ containing right-chiral fermions, additional gauge bosons, and a Higgs degree of freedom \cite{Kaushik2022E8}. The later split-bioctonion work supplies the geometric scaffolding for this picture by emphasizing the role of the two extra $SU(3)$ factors and the resulting $(3,3)$ backbone \cite{Singh2025SplitBioctonions}. Within the program, this is the structural reason the visible sector is not unified with gravity only at the Planck scale; the practically relevant matching point is the \ew\ scale, where the vacuum selects a triality orientation and the visible/right-handed split becomes phenomenologically operative \cite{Singh2023GraviWeak,Singh2025MassRatios}.

A second Higgs boson is therefore not an optional embellishment in this framework but part of the minimal structural consequence of the branching. So are right-handed gauge bosons and a dark photon after the right-handed sector breaks further. None of these features is unique to the program, but here they are not inserted ad hoc; they are meant to arise from the same exceptional organization that also generates the flavor relations.

\subsection{Sterile states, Majorana neutrinos, and leptonic CP}

The older literature already contained three right-handed sterile neutrinos \cite{Vaibhav2023Sterile}, but the more recent Jordan-algebra construction makes the neutrino claim more rigid. The 2022 analysis argued that the charged-sector mass-ratio construction sits naturally with Majorana neutrinos \cite{Bhatt2022Majorana}. The 2025 mass-ratio paper sharpens this into a stronger internal statement: within the present construction a Dirac neutrino is a no-go, whereas the Majorana option is selected, and the right-handed sector contains three inert Majorana states that are effectively invisible at low energy \cite{Singh2025MassRatios}.

This gives the neutrino package an unusual internal coherence: Majorana light neutrinos, inert right-handed Majorana partners, and --- as corrected in this version --- leptonic Dirac CP conservation, $\delta_{CP}^\ell\in\{0,\pi\}$ \cite{TeliSinghCP2026}. The corresponding v1 entry carried the maximal phase $\delta_{CP}^\ell=\pm\pi/2$, read off from the raw octonionic overlaps of \cite{Singh2025MassRatios}; the transport analysis of \cite{TeliSinghCP2026} shows that this reading arises from a removable global rephasing (a row-global factor of $i$), and proves that every admissible real transport yields a vanishing leptonic Jarlskog invariant, a physical phase being possible only if the Higgs bridge mixes the identity line with the lepton flavor plane. The corrected entry is sharper, not weaker: conditional, and precisely localized. Within normal ordering, CP-conserving values remain allowed by present global fits; within inverted ordering they are disfavored at roughly $3\sigma$ \cite{NuFIT2024} --- a tension the program accepts openly rather than hides. DUNE and Hyper-Kamiokande are the obvious experiments that can either strengthen or damage this part of the framework \cite{DUNE2022,HyperK2026}. For the Majorana claim itself, the decisive probes remain neutrinoless double-beta decay, complemented by cosmological and heavy-neutral-lepton constraints \cite{KamLANDZen2023,LEGEND2026,PDG2024}.

The 2026 companion papers make the neutrino sector substantially more committal. The leading right-handed Jordan spectrum $(-\delta_\nu,0,+\delta_\nu)$, read as the active Majorana texture under a stated minimality assumption, selects the \emph{inverted} mass ordering with a quasi-degenerate pair and one light state; unbroken dark electromagnetism forces any lift of this spectrum to be traceless, yielding the exact relation $m_2-m_1=m_3$ and hence, with oscillation data, $(m_1,m_2,m_3)=(49.07,\,49.84,\,0.762)$\,meV, $\Sigma m_\nu=0.0997$\,eV, and $m_{\beta\beta}=18.2\pm1.2$\,meV at the lower edge of the inverted-ordering band \cite{TeliSinghCP2026,SinghNu2026}. Independently of that assumption, the rank-2 structure of the Dirac block keeps the lightest neutrino massless in either ordering, so the cosmological sum is two-valued, $\Sigma m_\nu\in\{0.059,\,0.100\}$\,eV, with the quasi-degenerate continuum excluded \cite{SinghNu2026}.

The sterile sector has acquired a cosmological role. In the 2026 cosmology \cite{SinghCosmo2026} the three $\nu_R$ are diluted thermal relics that close the matter budget; the budget fixes only the trace of the sterile Majorana matrix, $\mathrm{tr}\,M_R\simeq113$\,eV ($\simeq40$\,eV per state if degenerate), with each mass free within a free-streaming window $\sim\![17,100]$\,eV, and the interface analysis \cite{SinghNu2026} verifies consistency with the fermion construction provided $M_R$ is trace-dominated (the minimal texture allowed by dark-charge conservation) and the active--sterile mixing is quarantined, $|U_{e4}|^2\lesssim10^{-6}$ --- a suppression that is structurally natural, since the neutrino carries neither of the two Abelian gradings and the Higgs bridge therefore has no rung to transport. Adopting the cosmological sterile sector eliminates the heavy-seesaw branch: the inverted-ordering package above then stands as the program's only neutrino option, and a JUNO normal-ordering determination would falsify the combined construction outright rather than reroute it. Sterile-exchange contributions to $0\nu\beta\beta$ are negligible, and KATRIN-type kink searches at $m_4\simeq40$\,eV are predicted null.

\subsection{Charged-fermion mass relations, CKM structure, and Koide shift}

The most distinctive quantitative sector of the program concerns charged-fermion masses. The mature presentation is now in the 2025 paper on fermion mass ratios \cite{Singh2025MassRatios}. The most familiar relation is the first-generation pattern
\begin{equation}
\sqrt{m_e}:\sqrt{m_u}:\sqrt{m_d}=1:2:3,
\qquad\text{equivalently}\qquad
m_e:m_u:m_d=1:4:9,
\end{equation}
intended to hold after a common-scale, common-scheme comparison near the \ew\ threshold.

This is not the whole story. The same framework predicts a Dynkin-swap relation between the down-quark and charged-lepton ladders,
\begin{equation}
\sqrt{\frac{m_\tau}{m_\mu}} = \sqrt{\frac{m_s}{m_d}},
\qquad\text{or equivalently}\qquad
\frac{m_\tau}{m_\mu} = \frac{m_s}{m_d},
\end{equation}
and also a structured set of up-sector adjacent ratios, CKM root-sum rules, and a small positive offset away from the exact Koide value after triality breaking \cite{Patel2023CKM,Singh2025MassRatios}. In the symmetric phase, the claim is stronger still: the framework says that an exact Koide-type relation is realized before triality breaking, and that the observed small post-breaking mismatch is itself part of the prediction rather than an embarrassment to be hidden \cite{Singh2025MassRatios,Koide1983}.

The conceptual importance of this sector is that the \SM\ predicts none of these algebraic relations. Yukawa matrices in the \SM\ accommodate masses and mixings; they do not explain them. If a single common-scale analysis were to show that the Jordan-ladder package fails in a systematic way, the framework would suffer real damage.

\subsection{Quantitative status of the sharpest claims}

A predictions paper earns trust by being more severe on its own sharpest outputs than an external critic would be. Table~\ref{tab:quant} therefore states, in one place, what the quantitatively sharp claims currently look like against the best available comparison points.

\begin{table}[p]
\centering
\footnotesize
\setlength{\tabcolsep}{3.5pt}
\renewcommand{\arraystretch}{0.96}
\caption{Current status of the sharpest quantitative claims. The purpose of the table is diagnostic, not promotional.}
\label{tab:quant}
\begin{tabularx}{\textwidth}{L{0.24\textwidth} L{0.22\textwidth} L{0.24\textwidth} Y}
\toprule
Claim & Theory output & Present comparison point & Status and caveat \\
\midrule
Low-energy electromagnetic coupling & $\alpha_{\rm em}^{\rm th}(0)=0.00729713629$, i.e. $\alpha^{-1}_{\rm th}(0)=137.04006$ \cite{Singh2026Gauge} & CODATA 2022 gives $\alpha^{-1}=137.035999177(21)$ \cite{Mohr2025} & Numerically close. This is one of the framework's cleanest outputs. \\
Strong coupling at $\MZ$ & $\alphas^{\rm th}(\MZ)=0.11675418$ \cite{Singh2026Gauge} & PDG average $\alphas(\MZ)=0.1180\pm0.0009$ \cite{PDG2024} & Also numerically close, though not exact. \\
Weak mixing angle & $\swt^{\rm th}=0.24969776$ \cite{Singh2026Gauge,Raj2022Bosonic} & Direct effective-angle measurements are near $0.2315$ \cite{LHCb2024WeakAngle} & Not presently a precision success. A scheme mismatch exists between $\swt$ and the effective leptonic angle, but even with that caveat the current comparison is visibly less impressive than for $\alpha$ and $\alphas$. \\
Mixed-regime coupling relation & $\alphas(\MZ)/\alphaem(0)=16$ exactly in the 2026 paper \cite{Singh2026Gauge} & Current mixed-regime proxy from PDG plus CODATA is numerically close to 16 & This should not be advertised as a same-scale renormalization-group identity. The 2026 derivation itself explicitly interprets it as a mixed-regime geometric matching. \\
Dynkin-swap charged-fermion relation & $\sqrt{m_\tau/m_\mu}=\sqrt{m_s/m_d}$, equivalently $m_\tau/m_\mu=m_s/m_d$ \cite{Singh2025MassRatios} & Singh's own first-pass \ew-scale check gives a square-root equality test of $0.9236$, stable in the range $0.922$--$0.924$ across scales \cite{Singh2025MassRatios} & Real tension is already visible: about a $7.5\%$ deficit from exact equality in the first-pass scan. The paper argues that one-time model-to-$\MSbar$ matching or finite corrections may still matter, but this is already a crisp discriminant. \\
First-generation relation & $\sqrt{m_e}:\sqrt{m_u}:\sqrt{m_d}=1:2:3$ \cite{Singh2025MassRatios} & Singh's own \MZ\ test gives $\sqrt{m_u}/(2\sqrt{m_e})\approx0.80$ and $\sqrt{m_d}/(3\sqrt{m_e})\approx0.78$ \cite{Singh2025MassRatios} & Again, there is visible tension: about $20\%$ low in the first pass. The framework now needs the promised model-specific matching calculation; otherwise this remains a real problem for the sharpest flavor relation. \\
Koide structure & Exact Koide before triality breaking; definite positive post-breaking offset \cite{Singh2025MassRatios} & Running-mass comparisons at the \ew\ scale are close to, but not exactly equal to, $2/3$ \cite{Singh2025MassRatios,HuangZhou2021} & Interesting and structured, but less decisive than the two mass-ratio tests above. \\
CKM: two of four degrees of freedom parameter-free & $|V_{us}|=0.2371$, $|V_{cb}|=0.0422$ (virtual-node transport, amplitude reading); $|V_{ub}|/|V_{cb}|=\sqrt{m_u/m_c}$ exact in the minimal model \cite{SinghCKM2026} & Global fit: $0.22501$, $0.04183$ & Deviations $+5.4\%$ and $+0.9\%$; the remaining two degrees of freedom ($\theta_{13}$, $\delta_{CP}$) reduce to one complex bridge element; fragilities (re-unitarization reading; $\sqrt{m_c/m_t}$ sensitivity, elasticity $-1.9$) disclosed. \\
Neutrino mass ordering & Inverted; $(m_1,m_2,m_3)=(49.07,49.84,0.762)$\,meV, $\Sigma m_\nu=0.0997$\,eV \cite{TeliSinghCP2026,SinghNu2026} & Global fits mildly prefer normal ordering; the minimal-$\Lambda$CDM DESI bound disfavors IO but does not transfer to the program's own cosmology \cite{NuFIT2024,SinghCosmo2026} & The program's sharpest dated exposure: JUNO decides, and in the combined construction a normal-ordering result falsifies outright. \\
Effective $0\nu\beta\beta$ mass & $m_{\beta\beta}=18.2\pm1.2$\,meV, opposite pair parities \cite{TeliSinghCP2026,SinghNu2026} & Current bounds $\sim28$--$122$\,meV \cite{KamLANDZen2023} & Inside LEGEND-1000/nEXO discovery reach; nuclear-matrix-element systematics blur the inverse inference. \\
Leptonic CP phase & $\delta_{CP}^\ell\in\{0,\pi\}$ unless the identity--flavor bridge component is nonzero \cite{TeliSinghCP2026} & Within IO, conservation disfavored at $\sim3\sigma$; within NO, allowed \cite{NuFIT2024} & Corrects the v1 maximal-phase entry; a measured phase localizes the bridge term rather than falsifying. \\
\bottomrule
\end{tabularx}
\end{table}

Three remarks follow from this table. First, the paper must no longer speak as if the mass-ratio sector were already an uncomplicated success. The framework's own 2025 analysis now openly records a $7.5\%$ deficit in the Dynkin-swap test and roughly $20\%$ deficits in the first-generation test before model-specific matching is performed \cite{Singh2025MassRatios}. Second, the mixed-regime relation $\alphas(\MZ)/\alphaem(0)=16$ must be presented with its own caveat: it is \emph{not} a same-scale renormalization-group identity, and should not be sold as one \cite{Singh2026Gauge}. Third, the weak-angle sector is currently the least persuasive of the dimensionless-coupling outputs and should be written in a more guarded register than the $\alpha$ and $\alphas$ results.

\subsection{Spin-1 Lorentz bosons and the non-fundamental graviton}

One further particle-physics prediction is often overlooked because it lies at the boundary with gravity. In the octonionic program, the deeper pre-quantum description contains universally interacting spin-1 Lorentz bosons rather than a fundamental spin-2 graviton \cite{Singh2021Alpha,Singh2026Emergent}. Classical four-dimensional gravity is then emergent after spontaneous localization of highly entangled fermionic states. This is not yet a direct collider observable, but it is a sharp structural divergence from perturbative quantum-gravity expectations and belongs in any honest catalogue of the framework's claims.

\section{Gravitation and cosmology}

\subsection{Dark electromagnetism as a new long-range force}

The gravitational and cosmological phenomenology of the program is organized around one central claim: there exists a new $U(1)$ interaction sourced not by mass itself but by a square-root mass label in the right-handed sector \cite{Finster2024DarkEM,Singh2026RelMOND}. This interaction, dubbed dark electromagnetism, is proposed as the microscopic origin of a relativistic MOND-like infrared regime. In the deep-MOND regime the force becomes effectively $1/r$, and the framework claims that this sector can reproduce the phenomenological successes usually associated with MOND while sitting inside the same exceptional unification story that also generates the particle-physics sector.

This is ambitious and testable. The program is not merely repeating the broad empirical statement that galaxy dynamics may deviate from Newtonian expectations below a critical acceleration; it is proposing a specific microscopic cause. That raises the evidential burden. The same framework must recover galaxy phenomenology where MOND works, reduce to general relativity in the high-acceleration regime, and survive cosmological and cluster-scale tests.

\subsection{Where the MOND-like sector is most exposed}

This is the sector where the framework is most vulnerable to near-term falsification, and the paper should say so plainly. Relativistic MOND theories are difficult. Bekenstein's TeVeS was a major milestone but also illustrated how hard it is to satisfy lensing and cosmological constraints simultaneously \cite{Bekenstein2004}. More recently, Skordis and Zlosnik constructed a relativistic MOND theory that can reproduce the cosmic microwave background and linear matter-power data without particle dark matter, showing that the relativistic MOND program is not dead -- but also that it requires substantial structure and careful tuning \cite{SkordisZlosnik2021}.

The \eights\ program has not yet demonstrated that its dark-electromagnetic realization clears the same hurdles. On clusters, MOND-like frameworks still typically require extra missing matter even when the discrepancy is reduced relative to Newtonian gravity with cold dark matter \cite{KelleherLelli2024}. On wide binaries, the observational picture remains actively contested. Banik \emph{et al.}\ reported strong constraints favoring Newtonian gravity over MOND in Gaia DR3 binaries \cite{Banik2024WideBinaries}; Pittordis, Sutherland, and Shepherd likewise found Newtonian models to fit better, while emphasizing that improved understanding of triple-system contamination remains important \cite{Pittordis2025WideBinaries}; and Cookson \emph{et al.}\ have now argued that, with stringent quality controls, there is no observational evidence for a MOND-induced velocity boost \cite{Cookson2026WideBinaries}. The original proposal to use wide binaries as a critical low-acceleration test of gravity goes back to Hernandez, Jim\'enez, and Allen, who argued that Solar-mass binaries with separations of order $7\times10^3\,\mathrm{au}$ enter the MOND regime and presented an initial Hipparcos/SDSS-based implementation of the test \cite{HernandezJimenezAllen2012}. Hernandez and collaborators later revisited the problem with Gaia DR2 and then with cleaner Gaia eDR3/DR3 samples, reporting Newtonian behaviour at smaller separations and an anomalous low-acceleration regime at larger separations, while repeatedly emphasizing the importance of sample purity, hidden tertiaries, projection effects, and the detailed construction of the control sample \cite{HernandezCortesAllenScarpa2019,HernandezCooksonCortes2022,Hernandez2023DR3,Hernandez2024Stats,Hernandez2024WideBinaryReview,HernandezKroupa2025}.

Independently, Chae developed a parallel sequence of analyses using different statistical pipelines. His 2023 Gaia DR3 study used an acceleration-plane deprojection method and argued for a breakdown of Newton--Einstein gravity below $g_N\sim 10^{-9}\,\mathrm{m\,s^{-2}}$ \cite{Chae2023Breakdown}. He then presented a cleaner ``statistically pure'' binary sample designed to suppress hidden companions \cite{Chae2024Robust}, a normalized-velocity-profile analysis with explicit Newtonian-versus-Milgromian model comparison \cite{Chae2024VelocityProfile}, and a Bayesian 3D orbit-modeling framework using Gaia DR3 radial velocities \cite{Chae2025Bayesian3D}. Most recently, Chae has extended this line of work to a more realistic orbit-by-orbit 3D inference scheme and a pilot HARPS-based sample with very precise radial velocities \cite{Chae2026HARPS}. Whether one agrees with these claims or not, the reader should see that the pro-anomaly side of the wide-binary literature is being advanced through several distinct pipelines rather than through a single paper or a single collaboration.
At the same time, Hernandez and collaborators have maintained that the methodology of several null analyses is itself problematic and have continued to argue for a low-acceleration anomaly in the wide-binary data \cite{Hernandez2024WideBinaryReview,HernandezKroupa2025}. The correct conclusion is therefore not that wide binaries already rule out any MOND-like infrared sector, but that this is now a sharply contested arena in which any new relativistic MOND mechanism must engage the literature directly rather than cite MOND successes in the abstract.

For the \eights\ program, that means the following. Dark electromagnetism should be presented as an interesting proposed mechanism for a relativistic MOND-like regime, but not yet as a phenomenologically established success. The open tasks are clear: cluster fits, lensing, wide-binary consistency, and CMB or large-scale-structure viability must all be shown in the concrete dark-electromagnetic realization itself rather than borrowed from the broader MOND literature.

\subsection{Parity-violating $SU(2)_R$ pre-gravitation and the gravi-weak link}

A distinctive theme of the recent literature is that gravitation and the weak interaction share a common origin. The 2023 Jordan-algebra paper argues that only right-handed particles participate in the pre-gravitational symmetry and that weak parity violation is a low-energy relic of that deeper asymmetry \cite{Singh2023GraviWeak}. The 2026 $SO(3,3)$ BF-theory paper sharpens this by giving one four-dimensional sector that reproduces Einstein gravity and a second opposite-signature sector whose symmetry breaking yields weak gauge dynamics \cite{Wesley2026SO33}.

The phenomenology here is still immature, but the structural claim is real. The weak interaction is not treated as just another internal gauge force: it is part of the same split-signature backbone that also supports the dark-electromagnetic sector and the proposed reinterpretation of nonlocality. This is why the framework repeatedly speaks of unifying the \SM\ with gravity at the \ew\ scale. In the program's own terms, the \ew\ transition is not merely when the Higgs gets a vacuum expectation value; it is when the deeper triality or pre-gravitational structure becomes oriented in the observed way.

\subsection{Semi-quantitative weak-scale estimates}

The gravi-weak literature also offers a semi-quantitative estimate of the Fermi scale and the weak-force range from geometric considerations related to the thickness of extra dimensions \cite{Singh2023GraviWeak}. These are not yet as sharp as the flavor relations, and they should not be advertised as finished precision predictions. But they do illustrate what the framework is trying to do: replace the standard hierarchy in which gravity enters only at the Planck scale with one in which the \ew\ scale is the relevant threshold for matching visible physics to pre-gravitation.

\subsection{Six-dimensional split-signature backbone and emergent gravity}

Several recent papers converge on a common geometric claim: the deeper arena of the theory is six-dimensional, of split signature $(3,3)$, and contains two overlapping Lorentzian sectors \cite{Furquan2025Timelike,Kaushik2022E8,Singh2025SplitBioctonions,Wesley2026SO33}. On its own, higher dimensionality is only a structural choice. It becomes predictive only when tied to concrete claims: Bell nonlocality, parity violation, dark electromagnetism, or a gravi-weak split.

The same geometric backbone is also used to motivate the program's nonstandard ontology of gravity. Classical gravity is emergent, not fundamental, and should not be quantized as if it were just another Yang--Mills field on a fixed background \cite{Singh2021Alpha,Singh2026Emergent}. Combined with the spin-1 Lorentz-boson sector, this means the microscopic theory is not expected to resemble a conventional graviton field theory. This is a meta-prediction rather than a direct low-energy observable, but it strongly constrains what the program can and cannot claim to be reproducing.

\section{Where can one obtain a truly sharp discriminant?}

The most important strategic question is not whether the program has testable claims in the broad sense. It does. The sharper question is whether any of those claims can become a discriminator comparable in spirit to Einstein's factor-of-two light bending. At present there are three serious candidates and one important secondary package.

\begin{table}[ht]
\centering
\caption{Candidate flagship tests, ordered by distinctiveness rather than maturity.}
\label{tab:flagship}
\begin{tabularx}{\textwidth}{L{0.26\textwidth} L{0.34\textwidth} Y}
\toprule
Candidate & Why it would be genuinely discriminatory & Main present obstacle \\
\midrule
Violation of the Tsirelson bound & Any loophole-free CHSH value above $2\sqrt{2}$ lies outside ordinary quantum mechanics, not merely outside the \SM & No derived magnitude: the programme's own assessment \cite{SinghTsirelson2026} declares the prediction not yet falsifiable until one exists; the proposed $10^{-5}$ null test, $S(\tau)$, and above-electroweak collider regimes define the search frontier \\
Fermion-only spontaneous collapse & Standard quantum theory predicts no intrinsic collapse at all, while generic collapse models do not naturally isolate fermions so sharply & A quantitative rate formula is still needed for realistic bosonic and fermionic mesoscopic systems \\
Global flavor and gauge overconstraint & The \SM\ predicts none of the Jordan-ladder mass relations, CKM rules, Koide shift, or mixed-regime coupling relation; the framework predicts several of them simultaneously & Less visually dramatic than a single discovery event and requires disciplined same-scheme, same-scale comparison \\
Neutrino--cosmology conjunction: inverted ordering $+$ $m_{\beta\beta}\simeq18$\,meV $+$ conditional CP conservation $+$ $\simeq\!40$\,eV cosmological steriles $+$ two-epoch $\Delta N_{\rm eff}$ & The conjunction is proprietary: no other framework predicts this correlated package, every element of which is committed in print before the data (Table~\ref{tab:ledger}) & Quark-mass and NME systematics; JUNO timeline $\sim$2031--32; modest $\Delta N_{\rm eff}^{\rm CMB}$ detection significance at CMB-S4 sensitivity \\
\bottomrule
\end{tabularx}
\end{table}

The first route remains the cleanest in principle. A convincing loophole-free Bell value above $2\sqrt{2}$ would be a direct discriminator between the \eights\ program and ordinary quantum mechanics. But until a concrete protocol exists, the paper should resist calling this a finished flagship prediction.

The second route is almost as striking. A genuine fermion-only collapse signal would distinguish the program not only from standard quantum mechanics but also from most collapse models currently on the market. Here again the next step is practical rather than rhetorical: a rate equation and an experimental design are worth more than another conceptual essay.

The third route may in fact be the nearest-term falsification path. In the present framework, the first-generation $1{:}4{:}9$ relation, the Dynkin-swap relation $m_\tau/m_\mu=m_s/m_d$, the CKM root-sum rules, the post-breaking Koide offset, the low-energy $\alpha$ value, and the relation $\alphas(\MZ)/\alphaem(0)=16$ are supposed to arise from one and the same octonionic or Jordan structure \cite{Singh2021Alpha,Patel2023CKM,Singh2025MassRatios,Singh2026Gauge}. A single well-controlled \ew-scale global analysis could therefore strongly support or strongly damage the entire construction without waiting for a new machine. The present paper's most concrete recommendation is therefore that such a global analysis should be written next.

\section{A predictions ledger}
\label{sec:ledger}

Table~\ref{tab:ledger} consolidates the program's sharpest commitments in pre-registration form: each entry lists where and when the claim was fixed, the experiment that decides it, and the condition under which it is dead. Entries are restricted to claims with a dated, realistic decision path; the quantum-foundational targets of Sec.~2 retain their own discussion. Rows 1, 2, 5, 6 and 7 constitute a single correlated package --- they stand or fall together, on a timeline that closes by roughly 2035.

{\small
\begin{longtable}{@{}L{0.28\textwidth} L{0.14\textwidth} L{0.17\textwidth} L{0.09\textwidth} L{0.22\textwidth}@{}}
\caption{Predictions ledger: claim, commitment, deciding experiment, timeline, kill condition.}\label{tab:ledger}\\
\toprule
Claim & Committed & Decided by & Timeline & Kill condition \\
\midrule
\endfirsthead
\toprule
Claim & Committed & Decided by & Timeline & Kill condition \\
\midrule
\endhead
1. Inverted neutrino mass ordering; no fallback in the combined construction & \cite{TeliSinghCP2026} (6/2026); \cite{SinghCosmo2026,SinghNu2026} (7/2026) & JUNO & $\sim$2031--32 & Normal-ordering determination falsifies the combined construction outright \\
2. $m_{\beta\beta}=18.2\pm1.2$\,meV & \cite{TeliSinghCP2026,SinghNu2026} & LEGEND-1000, nEXO & 2030s & Robust null across the inverted-ordering band (NME-aware) \\
3. $\Sigma m_\nu\in\{0.059,\,0.100\}$\,eV, nothing else; $0.0997$\,eV in the combined construction & \cite{SinghNu2026} & Cosmological $\Sigma m_\nu$, analysed within the program's own background & late 2020s--2030s & $\Sigma$ robustly at neither value \\
4. $m_3=0.762\pm0.011$\,meV via the exact lift relation $m_2-m_1=m_3$ & \cite{SinghNu2026} & Internal consistency with precision solar/reactor data & ongoing & Violation of the traceless-lift relation \\
5. Two-epoch dark radiation: $\Delta N_{\rm eff}^{\rm BBN}\simeq0.19$, $\Delta N_{\rm eff}^{\rm CMB}\simeq0.054$ & \cite{SinghCosmo2026} & Simons Observatory, CMB-S4; D/H and $Y_p$ & late 2020s--2030s & Robust $\Delta N_{\rm eff}^{\rm CMB}=0$ at $\sigma\simeq0.03$, or equal BBN/CMB values \\
6. Sterile sector: $\mathrm{tr}\,M_R\simeq113$\,eV, each mass in $\sim[17,100]$\,eV; KATRIN kink null, $|U_{e4}|^2\lesssim10^{-6}$ & \cite{SinghCosmo2026,SinghNu2026} & KATRIN/TRISTAN; $P(k)$ free-streaming split & now--2030s & Kink detection at $|U_{e4}|^2\gtrsim10^{-3}$; sterile clustering on galactic scales \\
7. Permanent nulls: no WIMP, no collider dark matter, no fourth generation & \cite{SinghCosmo2026} and earlier & LZ/XLZD, LHC/FCC & permanent & Any confirmed particle-dark-matter or fourth-generation detection \\
8. $\delta_{CP}^\ell\in\{0,\pi\}$ (conditional on a real Higgs bridge) & \cite{TeliSinghCP2026} & DUNE, Hyper-Kamiokande & 2030s & Not a kill: a measured phase localizes the bridge term (recorded for honesty) \\
9. $\sqrt{m_\tau/m_\mu}=\sqrt{m_s/m_d}$ & \cite{Singh2025MassRatios} & Lattice $m_s/m_d$ at a common \ew\ scale & ongoing & $>3\sigma$ exclusion of equality \\
10. $\alphas(\MZ)/\alphaem(0)=16$ & \cite{Singh2026Gauge} & Existing precision data, scheme-consistent analysis & now & Scheme-consistent analysis contradicts the relation \\
11. Gravitational parity: everyday gravity parity-blind (screened); any chirality/CP asymmetry of the gravitational coupling below $3\times10^{-9}$ & \cite{SinghParity2026} & Galactic-supernova triple timing ($\nu_e$, $\bar\nu_e$, GW) & next Galactic SN & Detected asymmetry falsifies the screening structure \\
12. Chiral graviton vacuum ($\gamma=\pm i$): nonzero CMB TB/EB correlations, conditional on $r\gtrsim\text{few}\times10^{-3}$ & \cite{SinghParity2026} & LiteBIRD, CMB-S4 & 2030s & $r$ detected with TB/EB absent at the predicted level \\
13. CKM internal correlations: $|V_{cb}|$ tied to $\sqrt{m_t/m_c}$ (elasticity $-1.9$) and $|V_{us}|$ to $\sqrt{m_s/m_d}$; parameter-free values $0.0422$, $0.2371$ & \cite{SinghCKM2026} & Lattice and global fits; the programme's own matching refinements & ongoing & Correlated motion violated; or global fit departs from $0.0422$ beyond the disclosed fragility band \\
\bottomrule
\end{longtable}}

The commitment dates are verifiable from the public version histories of the cited preprints. We highlight the asymmetry deliberately: several rows are falsification-decisive but confirmation-shared (inverted ordering, for instance, is predicted by other models too); what is proprietary is the conjunction, together with the exposure --- row 1 in particular admits no retreat.

\section{Conclusions}

The \eights\ octonionic program already makes a wider and sharper set of empirical claims than a casual summary would suggest. In quantum foundations these include objective spontaneous collapse, operator time, spontaneous collapse in time, an attosecond-scale temporal-interference cutoff, a six-dimensional explanation of apparent nonlocality, a possible window above the Tsirelson bound, a fermion-only collapse sector, and holographic or Karolyhazy uncertainty. In particle physics the mature list includes an extended Higgs sector, a right-handed gauge sector, a dark photon, three right-handed Majorana neutrinos (now the $\simeq\!40$\,eV cosmological relics), Majorana light neutrinos, inverted mass ordering with $m_{\beta\beta}\simeq18$\,meV, conditional leptonic CP conservation, CKM root-sum rules, several parameter-free charged-fermion mass relations including $m_\tau/m_\mu=m_s/m_d$ and the first-generation $1{:}4{:}9$ pattern, the low-energy fine-structure constant, and the mixed-regime relation $\alphas(\MZ)/\alphaem(0)=16$. In gravitation and cosmology the key claims are dark electromagnetism, a relativistic MOND-like infrared regime with $a_0$ anchored to $\Lambda$, the two-epoch dark-radiation fingerprint, parity-violating $SU(2)_R$ pre-gravitation, a six-dimensional gravi-weak backbone, and emergent classical gravity.

At the same time, the paper should no longer read as if all these predictions stand on equal footing or all are already successes. The flavor sector is the most algebraically concrete, but it now openly carries quantitative tensions that must be confronted rather than ignored. The gauge-coupling sector contains numerically attractive outputs, but the mixed-regime nature of the $\alphas/\alphaem$ relation must be stated explicitly. The weak-angle derivation is presently the least convincing of the sharp coupling claims. The neutrino package is coherent but awaits decisive data. The dark-electromagnetic and gravi-weak sectors are bold, but they still require much harder phenomenological implementation against the current literature on relativistic MOND, clusters, wide binaries, and cosmology.

In v1 this paper stated plainly that the program still lacked a universally accepted Einstein-style discriminator. The 2026 developments have changed that assessment. The neutrino--cosmology conjunction of Table~\ref{tab:ledger} --- inverted ordering, $m_{\beta\beta}\simeq18$\,meV, the two-epoch dark-radiation pattern, the quarantined $40$\,eV sterile sector, and permanent dark-matter-search nulls, committed in print before the data and, in the combined construction, with no fallback branch --- now functions as the program's eclipse bet, with JUNO's ordering determination playing the role of the eclipse. The complementary routes remain valuable: derive a concrete Bell protocol that produces a calculable super-quantum window, turn fermion-only collapse into a quantitative differential test, and perform a fully scheme-consistent \ew-scale global analysis of the intertwined mass, mixing, and coupling relations. A program that claims to unify quantum foundations, particle physics, and gravity should eventually live or die by tests of that severity.

\paragraph{Note added (v2).} Three changes relative to v1 (April 2026). First, the leptonic CP entry is corrected: the maximal phase $\delta_{CP}^\ell=\pm\pi/2$ carried by v1 originated in a removable rephasing of the raw octonionic overlaps, and is replaced by the transport-theorem result $\delta_{CP}^\ell\in\{0,\pi\}$, conditional and localized \cite{TeliSinghCP2026}. Second, the sterile-neutrino sector is updated to its cosmological role (three diluted $\simeq\!40$\,eV relics closing the matter budget), with the ordering, $m_{\beta\beta}$, $\Sigma m_\nu$ and dark-radiation entries updated accordingly \cite{SinghCosmo2026,SinghNu2026}. Third, the predictions ledger of Sec.~\ref{sec:ledger} is added. Fourth, the second-Higgs, gravitational-parity and CKM entries are upgraded to the quantitative profiles of the dedicated 2026 papers \cite{SinghHiggs2026,SinghParity2026,SinghCKM2026}: the second Higgs becomes a full electroweak quartet with a search strategy; the parity of gravity becomes a derived screening statement with a named residual operator basket and two new ledger rows; and the CKM sector acquires two parameter-free degrees of freedom with disclosed fragilities. The Tsirelson entry is updated to the dedicated 2026 assessment \cite{SinghTsirelson2026}: mechanism and sign supplied, magnitude not yet derived, with the experimental record and five proposed tests recorded. The catalogue structure is otherwise unchanged.

\section*{Acknowledgements}

\noindent{\bf Use of generative AI}: This document has been prepared with assistance from Anthropic's Claude Fable 5. The author takes full intellectual responsibility for the contents of the article.


\begin{thebibliography}{99}

\bibitem{Singh2018QTSTS}
T. P. Singh,
``Quantum theory and the structure of space-time,''
\textit{Z. Naturforsch. A} \textbf{73}, 733 (2018).

\bibitem{Singh2018GRWST}
T. P. Singh,
``Space and time as a consequence of GRW quantum jumps,''
\textit{Z. Naturforsch. A} \textbf{73}, 923 (2018).

\bibitem{Singh2019STCollapse}
T. P. Singh,
``Space-time from collapse of the wave-function,''
\textit{Z. Naturforsch. A} \textbf{74}, 147 (2019).

\bibitem{Adler2004}
S. L. Adler,
\textit{Quantum Theory as an Emergent Phenomenon}
(Cambridge University Press, Cambridge, 2004).

\bibitem{Singh2021Alpha}
T. P. Singh,
``Quantum theory without classical time: octonions, and a theoretical derivation of the fine structure constant 1/137,''
\textit{Int. J. Mod. Phys. D} \textbf{30}, 2142010 (2021).

\bibitem{Singh:2022lpw}
T.~P.~Singh,
``Quantum gravity effects in the infrared: a theoretical derivation of the low-energy fine structure constant and mass ratios of elementary particles,''
Eur. Phys. J. Plus \textbf{137} (2022) no.6, 664
doi:10.1140/epjp/s13360-022-02868-4
[arXiv:2205.06614 [physics.gen-ph]].

\bibitem{Kaushik2022E8}
P. Kaushik, V. Vaibhav, and T. P. Singh,
``An $E_8\times E_8$ unification of the standard model with pre-gravitation, on an exceptional Lie algebra-valued space,''
arXiv:2206.06911 [hep-ph].

\bibitem{Singh2025SplitBioctonions}
T. P. Singh,
``Spacetime and internal symmetry from split bioctonions and the two extra $SU(3)$'s of $E_8\times E_8$,''
Preprints \textbf{2025}, 2025100437 (2025).

\bibitem{Singh2026Emergent}
T.~P.~Singh,
``Gravitation, and quantum theory, as emergent phenomena,''
J. Phys. Conf. Ser. \textbf{2533} (2023) no.1, 012013
doi:10.1088/1742-6596/2533/1/012013
[arXiv:2308.16216 [physics.gen-ph]].


\bibitem{Ahmed2022Tsirelson}
R. G. Ahmed and T. P. Singh,
``A violation of the Tsirelson bound in the pre-quantum theory of trace dynamics,''
arXiv:2208.02209 [quant-ph].

\bibitem{Furquan2025Timelike}
M. Furquan, T. P. Singh, and P. S. Wesley,
``Time-like extra dimensions: non-locality, spin, and Tsirelson bound,''
\textit{Universe} \textbf{11}, 137 (2025).

\bibitem{Cirelson1980}
B. S. Cirel'son,
``Quantum generalizations of Bell's inequality,''
\textit{Lett. Math. Phys.} \textbf{4}, 93 (1980).

\bibitem{PopescuRohrlich1994}
S. Popescu and D. Rohrlich,
``Quantum nonlocality as an axiom,''
\textit{Found. Phys.} \textbf{24}, 379 (1994).

\bibitem{Singh2026FermionCollapse}
T. P. Singh,
``In models of spontaneous wave-function collapse, why only fermions collapse, not bosons?,''
arXiv:2602.15044 [quant-ph].

\bibitem{Singh2025MassRatios}
T. P. Singh,
``Fermion mass ratios from the exceptional Jordan algebra,''
arXiv:2508.10131 [hep-ph].

\bibitem{Singh2026Gauge}
T. P. Singh, "Gauge Couplings of the Standard Model in the octonionic framework: a broken phase mechanism for $\alpha_s/\alpha_{em}=16$", arXiv:2603.28810 [hep-ph]

\bibitem{Jordan1934}
P. Jordan, J. von Neumann, and E. Wigner,
``On an algebraic generalization of the quantum mechanical formalism,''
\textit{Ann. Math.} \textbf{35}, 29 (1934).

\bibitem{GunaydinGursey1973}
M. G\"unaydin and F. G\"ursey,
``Quark structure and octonions,''
\textit{J. Math. Phys.} \textbf{14}, 1651 (1973).

\bibitem{GunaydinGursey1974}
M. G\"unaydin and F. G\"ursey,
``Quark statistics and octonions,''
\textit{Phys. Rev. D} \textbf{9}, 3387 (1974).

\bibitem{Baez2002}
J. C. Baez,
``The octonions,''
\textit{Bull. Amer. Math. Soc.} \textbf{39}, 145 (2002).

\bibitem{Furey2014}
C. Furey,
``Generations: three prints, in colour,''
\textit{JHEP} \textbf{10}, 046 (2014).

\bibitem{Furey2018}
C. Furey,
``Three generations, two unbroken gauge symmetries, and one eight-dimensional algebra,''
\textit{Phys. Lett. B} \textbf{785}, 84 (2018).

\bibitem{GillardGresnigt2019}
A. B. Gillard and N. G. Gresnigt,
``Three fermion generations with two unbroken gauge symmetries from the complex sedenions,''
\textit{Eur. Phys. J. C} \textbf{79}, 446 (2019).

\bibitem{Todorov2023}
I. Todorov,
``Octonion internal space algebra for the Standard Model,''
\textit{Universe} \textbf{9}, 222 (2023).

\bibitem{Bhatt2022Majorana}
V. Bhatt, R. Mondal, V. Vaibhav, and T. P. Singh,
``Majorana neutrinos, exceptional Jordan algebra, and mass ratios for charged fermions,''
\textit{J. Phys. G} \textbf{49}, 045007 (2022).

\bibitem{Patel2023CKM}
A. A. Patel and T. P. Singh,
``CKM matrix parameters from the exceptional Jordan algebra,''
\textit{Universe} \textbf{9}, 440 (2023).

\bibitem{Singh2023GraviWeak}
T. P. Singh,
``The exceptional Jordan algebra, and its implications for our understanding of gravitation and the weak force,''
arXiv:2304.01213 [gr-qc].

\bibitem{Wesley2026SO33}
P. S. Wesley, T. P. Singh, and J. M. Isidro,
``Gravity and electroweak sector from symmetry breaking of an $SO(3,3)$ BF theory,''
arXiv:2602.19151 [hep-th].

\bibitem{Finster2024DarkEM}
F. Finster, J. M. Isidro, C. F. Paganini, and T. P. Singh,
``Theoretically motivated dark electromagnetism as the origin of relativistic MOND,''
\textit{Universe} \textbf{10}, 123 (2024).

\bibitem{Singh2026RelMOND}
T. P. Singh,
``A relativistic MOND,''
arXiv:2601.04290 [gr-qc].

\bibitem{HuangZhou2021}
G.-Y. Huang and S. Zhou,
``Precise values of running quark and lepton masses in the standard model,''
\textit{Phys. Rev. D} \textbf{103}, 016010 (2021).

\bibitem{Mishra2018BulkHeat}
R. Mishra, A. Vinante, and T. P. Singh,
``Testing spontaneous collapse through bulk heating experiments: estimate of the background noise,''
\textit{Phys. Rev. A} \textbf{98}, 052121 (2018).

\bibitem{GRW1986}
G. C. Ghirardi, A. Rimini, and T. Weber,
``Unified dynamics for microscopic and macroscopic systems,''
\textit{Phys. Rev. D} \textbf{34}, 470 (1986).

\bibitem{Pearle1989}
P. Pearle,
``Combining stochastic dynamical state-vector reduction with spontaneous localization,''
\textit{Phys. Rev. A} \textbf{39}, 2277 (1989).

\bibitem{Bassi2013}
A. Bassi, K. Lochan, S. Satin, T. P. Singh, and H. Ulbricht,
``Models of wave-function collapse, underlying theories, and experimental tests,''
\textit{Rev. Mod. Phys.} \textbf{85}, 471 (2013).

\bibitem{Bassi2023}
A. Bassi, M. Dorato, and H. Ulbricht,
``Collapse models: a theoretical, experimental and philosophical review,''
\textit{Entropy} \textbf{25}, 645 (2023).

\bibitem{Carlesso:2022pqr}
M.~Carlesso, S.~Donadi, L.~Ferialdi, M.~Paternostro, H.~Ulbricht and A.~Bassi,
``Present status and future challenges of non-interferometric tests of collapse models,''
Nature Phys. \textbf{18} (2022) no.3, 243-250
doi:10.1038/s41567-021-01489-5
[arXiv:2203.04231 [quant-ph]].

\bibitem{Aprile2026XENON}
E. Aprile \textit{et al.} (XENON Collaboration),
``Challenging spontaneous quantum collapse with XENONnT,''
\textit{Phys. Rev. Lett.} \textbf{136}, 120201 (2026).


\bibitem{Kakade2023Spont}
K. Kakade, A. Singh, and T. P. Singh,
``Spontaneous localisation from a coarse-grained deterministic and non-unitary dynamics,''
\textit{Phys. Lett. A} \textbf{490}, 129191 (2023).



\bibitem{Lindner2005}
F. Lindner \textit{et al.},
``Attosecond double-slit experiment,''
\textit{Phys. Rev. Lett.} \textbf{95}, 040401 (2005).

\bibitem{Kaneyasu2023}
T. Kaneyasu \textit{et al.},
``Time domain double slit interference of electron produced by XUV synchrotron radiation,''
\textit{Sci. Rep.} \textbf{13}, 6142 (2023).

\bibitem{JavedWilsonEwing2026}
I. Javed and E. Wilson-Ewing,
``Testing wormhole-mediated entanglement with hydrogen,''
\textit{Phys. Rev. Lett.} \textbf{136}, 121501 (2026).

\bibitem{Hensen2015}
B. Hensen \textit{et al.},
``Loophole-free Bell inequality violation using electron spins separated by 1.3 kilometres,''
\textit{Nature} \textbf{526}, 682 (2015).

\bibitem{Giustina2015}
M. Giustina \textit{et al.},
``Significant-loophole-free test of Bell's theorem with entangled photons,''
\textit{Phys. Rev. Lett.} \textbf{115}, 250401 (2015).

\bibitem{Shalm2015}
L. K. Shalm \textit{et al.},
``Strong loophole-free test of local realism,''
\textit{Phys. Rev. Lett.} \textbf{115}, 250402 (2015).

\bibitem{Singh2021Holography}
T. P. Singh,
``Quantum gravity, holography, and minimal length,''
\textit{Pramana} \textbf{95}, 40 (2021).

\bibitem{Karolyhazy1966}
F. Karolyhazy,
``Gravitation and quantum mechanics of macroscopic objects,''
\textit{Nuovo Cimento A} \textbf{42}, 390 (1966).

\bibitem{NgVanDam1994}
Y. J. Ng and H. van Dam,
``Limit to space-time measurement,''
\textit{Mod. Phys. Lett. A} \textbf{9}, 335 (1994).

\bibitem{Chou2017}
A. Chou \textit{et al.},
``Interferometric constraints on quantum geometrical shear noise correlations,''
\textit{Class. Quantum Grav.} \textbf{34}, 165005 (2017).

\bibitem{Richardson2021}
J. W. Richardson \textit{et al.},
``Interferometric constraints on spacelike coherent rotational fluctuations,''
\textit{Phys. Rev. Lett.} \textbf{126}, 241301 (2021).

\bibitem{PDG2024}
S. Navas \textit{et al.} (Particle Data Group),
``Review of Particle Physics,''
\textit{Phys. Rev. D} \textbf{110}, 030001 (2024).

\bibitem{Vaibhav2023Sterile}
V. Vaibhav and T. P. Singh,
``Left-right symmetric fermions and sterile neutrinos from complex split biquaternions and bioctonions,''
\textit{Adv. Appl. Clifford Algebras} \textbf{33}, 32 (2023).

\bibitem{KamLANDZen2023}
S. Abe \textit{et al.} (KamLAND-Zen Collaboration),
``Search for the Majorana nature of neutrinos in the inverted mass ordering region with KamLAND-Zen,''
\textit{Phys. Rev. Lett.} \textbf{130}, 051801 (2023).

\bibitem{LEGEND2026}
H. Acharya \textit{et al.} (LEGEND Collaboration),
``First results on the search for lepton number violating neutrinoless double-$\beta$ decay with the LEGEND-200 experiment,''
\textit{Phys. Rev. Lett.} \textbf{136}, 022701 (2026).

\bibitem{NuFIT2024}
I. Esteban, M. C. Gonzalez-Garcia, M. Maltoni, I. Martinez-Soler, J. P. Pinheiro, and T. Schwetz,
``NuFIT-6.0: updated global analysis of three-flavor neutrino oscillations,''
\textit{JHEP} \textbf{12}, 216 (2024).

\bibitem{DUNE2022}
DUNE Collaboration,
``Low exposure long-baseline neutrino oscillation sensitivity of the DUNE experiment,''
\textit{Phys. Rev. D} \textbf{105}, 072006 (2022).

\bibitem{HyperK2026}
Hyper-Kamiokande Collaboration,
``Sensitivity of the Hyper-Kamiokande experiment to neutrino oscillation parameters using accelerator neutrinos,''
\textit{Eur. Phys. J. C} \textbf{86}, 170 (2026).

\bibitem{Koide1983}
Y. Koide,
``New view of quark and lepton mass hierarchy,''
\textit{Phys. Rev. D} \textbf{28}, 252 (1983).

\bibitem{Mohr2025}
P. J. Mohr, D. B. Newell, B. N. Taylor, and E. Tiesinga,
``CODATA recommended values of the fundamental physical constants: 2022,''
\textit{Rev. Mod. Phys.} \textbf{97}, 025002 (2025).

\bibitem{Raj2022Bosonic}
S. Raj and T. P. Singh,
``A Lagrangian with $E_8\times E_8$ symmetry for the standard model and pre-gravitation I: the bosonic Lagrangian, and a theoretical derivation of the weak mixing angle,''
arXiv:2208.09811 [hep-ph].

\bibitem{LHCb2024WeakAngle}
R. Aaij \textit{et al.} (LHCb Collaboration),
``Measurement of the effective leptonic weak mixing angle,''
\textit{JHEP} \textbf{12}, 026 (2024).

\bibitem{Milgrom1983}
M. Milgrom,
``A modification of the Newtonian dynamics as a possible alternative to the hidden mass hypothesis,''
\textit{Astrophys. J.} \textbf{270}, 365 (1983).

\bibitem{Bekenstein2004}
J. D. Bekenstein,
``Relativistic gravitation theory for the MOND paradigm,''
\textit{Phys. Rev. D} \textbf{70}, 083509 (2004).

\bibitem{SkordisZlosnik2021}
C. Skordis and T. Zlosnik,
``A new relativistic theory for modified Newtonian dynamics,''
\textit{Phys. Rev. Lett.} \textbf{127}, 161302 (2021).

\bibitem{Bell1964}
J. S. Bell,
``On the Einstein Podolsky Rosen paradox,''
\textit{Physics} \textbf{1}, 195 (1964).

\bibitem{CHSH1969}
J. F. Clauser, M. A. Horne, A. Shimony, and R. A. Holt,
``Proposed experiment to test local hidden-variable theories,''
\textit{Phys. Rev. Lett.} \textbf{23}, 880 (1969).

\bibitem{KelleherLelli2024}
R. Kelleher and F. Lelli,
``Galaxy clusters in Milgromian dynamics: missing matter, hydrostatic bias, and the external field effect,''
\textit{Astron. Astrophys.} \textbf{688}, A78 (2024).

\bibitem{Banik2024WideBinaries}
I. Banik, C. Pittordis, W. Sutherland, B. Famaey, R. Ibata, S. Mieske, and H. Zhao,
``Strong constraints on the gravitational law from Gaia DR3 wide binaries,''
\textit{Mon. Not. R. Astron. Soc.} \textbf{527}, 4573 (2024).

\bibitem{Pittordis2025WideBinaries}
C. Pittordis, W. Sutherland, and P. Shepherd,
``Wide binaries from Gaia DR3: testing GR vs MOND with realistic triple modelling,''
\textit{Open J. Astrophys.} \textbf{8}, 109 (2025).

\bibitem{Cookson2026WideBinaries}
S. A. Cookson, I. Banik, K. El-Badry, and C. J. Clarke,
``A quality framework for testing gravity with wide binaries: no evidence for MOND,''
\textit{Mon. Not. R. Astron. Soc.} \textbf{547}, stag342 (2026).

\bibitem{HernandezJimenezAllen2012}
X.~Hernandez, M.~A.~Jim\'enez, and C.~Allen,
``Wide binaries as a critical test of classical gravity,''
Eur.\ Phys.\ J.\ C \textbf{72}, 1884 (2012),
doi:10.1140/epjc/s10052-012-1884-6.

\bibitem{HernandezCortesAllenScarpa2019}
X.~Hernandez, R.~A.~M.~Cort\'es, C.~Allen, and R.~Scarpa,
``Challenging a Newtonian prediction through Gaia wide binaries,''
Int.\ J.\ Mod.\ Phys.\ D \textbf{28}, 1950101 (2019),
doi:10.1142/S0218271819501013.

\bibitem{HernandezCooksonCortes2022}
X.~Hernandez, S.~Cookson, and R.~A.~M.~Cort\'es,
``Internal kinematics of Gaia eDR3 wide binaries,''
Mon.\ Not.\ R.\ Astron.\ Soc.\ \textbf{509}, 2304--2317 (2022),
doi:10.1093/mnras/stab3038.

\bibitem{Hernandez2023DR3}
X.~Hernandez,
``Internal kinematics of Gaia DR3 wide binaries: anomalous behaviour in the low acceleration regime,''
Mon.\ Not.\ R.\ Astron.\ Soc.\ \textbf{525}, 1401--1415 (2023),
doi:10.1093/mnras/stad2306.

\bibitem{Hernandez2024Stats}
X.~Hernandez, V.~Verteletskyi, L.~Nasser, and A.~Aguayo-Ortiz,
``Statistical analysis of the gravitational anomaly in Gaia wide binaries,''
Mon.\ Not.\ R.\ Astron.\ Soc.\ \textbf{528}, 4720--4732 (2024),
doi:10.1093/mnras/stad3446.


\bibitem{Hernandez2024WideBinaryReview}
X. Hernandez and K.-H. Chae,
``A critical review of recent Gaia wide binary gravity tests,''
\textit{Mon. Not. R. Astron. Soc.} \textbf{533}, 729 (2024).

\bibitem{HernandezKroupa2025}
X. Hernandez and P. Kroupa,
``A recent confirmation of the wide binary gravitational anomaly,''
\textit{Mon. Not. R. Astron. Soc.} \textbf{537}, 2925 (2025).


\bibitem{Chae2023Breakdown}
K.-H.~Chae,
``Breakdown of the Newton--Einstein standard gravity at low acceleration in internal dynamics of wide binary stars,''
Astrophys.\ J.\ \textbf{952}, 128 (2023),
doi:10.3847/1538-4357/ace101.

\bibitem{Chae2024Robust}
K.-H.~Chae,
``Robust evidence for the breakdown of standard gravity at low acceleration from statistically pure binaries free of hidden companions,''
Astrophys.\ J.\ \textbf{960}, 114 (2024),
doi:10.3847/1538-4357/ad0ed5.

\bibitem{Chae2024VelocityProfile}
K.-H.~Chae,
``Measurements of the low-acceleration gravitational anomaly from the normalized velocity profile of Gaia wide binary stars and statistical testing of Newtonian and Milgromian theories,''
Astrophys.\ J.\ \textbf{972}, 186 (2024),
doi:10.3847/1538-4357/ad61e9.

\bibitem{Chae2025Bayesian3D}
K.-H.~Chae,
``Low-acceleration gravitational anomaly from Bayesian 3D modeling of wide binary orbits: methodology and results with Gaia Data Release 3,''
Astrophys.\ J.\ \textbf{985}, 210 (2025),
doi:10.3847/1538-4357/adce09.

\bibitem{Chae2026HARPS}
K.-H.~Chae,
``Bayesian inference of gravity through realistic 3D modeling of wide binary orbits: general algorithm and a pilot study with HARPS radial velocities,''
Astrophys.\ J.\ Lett.\ \textbf{998}, L43 (2026),
doi:10.3847/2041-8213/ae40ef.
\bibitem{TeliSinghCP2026}
B. G. Teli and T. P. Singh,
``Leptonic CP conservation and the quark CP phase from octonionic flavor structure,''
arXiv:2606.27836 [hep-ph] (2026).

\bibitem{TeliSinghMass2026}
B. G. Teli and T. P. Singh,
``Fermion mass hierarchies and the exceptional Jordan algebra,''
arXiv:2605.24866 [hep-ph] (2026).

\bibitem{TeliSinghMix2026}
B. G. Teli and T. P. Singh,
``Fermion mixing matrices and the exceptional Jordan algebra,''
arXiv:2607.00412 [hep-ph] (2026).

\bibitem{SinghCosmo2026}
T. P. Singh,
``A cosmological sector for the $E_8\times\omega E_8$ octonionic unification: emergence at the electroweak epoch, infrared gravity in place of dark matter, and a tilted late-time universe,''
preprint (2026), doi:10.5281/zenodo.21195649.

\bibitem{SinghNu2026}
T. P. Singh,
``The neutrino sector of the $E_8\times\omega E_8$ octonionic programme: bridge triage, degeneracy lifting, and a TeV-scale seesaw under the desert,''
preprint (2026), doi:10.5281/zenodo.21196194.

\bibitem{SinghHiggs2026}
T. P. Singh,
``The second Higgs of the $E_8\times\omega E_8$ octonionic unification program: quantum numbers, effective couplings, and search strategy,''
preprint (2026), doi:10.5281/zenodo.21196808.

\bibitem{SinghParity2026}
T. P. Singh,
``Does gravity violate parity? Source currents, screening theorems, and experimental tests of $SU(2)_R$ pre-gravitation,''
preprint (2026), doi:10.5281/zenodo.21194449.

\bibitem{SinghTsirelson2026}
T. P. Singh,
``Prospects for an experimental test of the Tsirelson-bound violation predicted by the $E_8\times\omega E_8$ octonionic--trace-dynamics unification programme,''
preprint (2026), doi:10.5281/zenodo.21194041.

\bibitem{SinghCKM2026}
T. P. Singh,
``The CKM sector of the exceptional-Jordan programme, reduced to its four degrees of freedom,''
preprint (2026), doi:10.5281/zenodo.21196380.

\bibitem{JUNO2024Order}
JUNO Collaboration,
``Potential to identify the neutrino mass ordering with reactor antineutrinos at JUNO,''
Chin. Phys. C, arXiv:2405.18008 (2024); first oscillation results: arXiv:2511.14593 (2025).

\end{thebibliography}
\end{document}